\documentclass[a4paper,10pt,twoside]{cpc-hepnp}

\usepackage{multicol}
\usepackage{graphicx}
\usepackage{booktabs}
\usepackage{amssymb,bm,mathrsfs,bbm,amscd}
\usepackage[tbtags]{amsmath}
\usepackage{lastpage}

\begin{document}
%\begin{CJK*}{GBK}{song}

\fancyhead[c]{\small Chinese Physics C~~~Vol. **, No. * (****)
******} \fancyfoot[C]{\small ******-\thepage}

\footnotetext[0]{Received ** ***** ****}

\title{The influence of the Coulomb exchange term on nuclear single-proton resonances\thanks{Supported by National Natural Science Foundation of China (11205004) }}

\author{%
      WANG Shu-Yang$^{1}$
\quad ZHU Zhong-Lai$^{1}$
\quad NIU Zhong-Ming$^{1;1)}$\email{zmniu@ahu.edu.cn}%
}
\maketitle

\address{%
$^1$ School of Physics and Material Science, Anhui University, Hefei 230601, China\\
}

\begin{abstract}
Nuclear single-proton resonances are sensitive to the Coulomb field, while the exchange term of Coulomb field is usually neglected due to its nonlocality. By combining the complex scaling method with the relativistic mean-field model, the influence of the Coulomb exchange term on the single-proton resonances is investigated by taking Sn isotopes and $N=82$ isotones as examples. It is found that the Coulomb exchange term reduces the single-proton resonance energy within the range of $0.4-0.6$ MeV, and lead to similar isotopic and isotonic trends of the resonance energy as those without the Coulomb exchange term. Moreover, the single-proton resonance width is also reduced by the Coulomb exchange term, whose influence generally decreases with the increasing neutron number and increases with the increasing proton number. However, the influence of the Coulomb exchange term cannot change the trend of the resonance width with respect to the neutron number and proton number. Furthermore, the influence of the Coulomb exchange term on the resonance width is investigated for the doubly magic nuclei $^{40}$Ca, $^{56,78}$Ni, $^{100,132}$Sn, and $^{208}$Pb. It is found that the Coulomb exchange term reduces the proton resonance width within $0.2$ MeV, whose magnitude depends on the specific nucleus and the quantum numbers of resonant states.
\end{abstract}



\footnotetext[0]{\hspace*{-3mm}\raisebox{0.3ex}{$\scriptstyle\copyright$}2013
Chinese Physical Society and the Institute of High Energy Physics of the Chinese Academy of Sciences and the Institute of Modern Physics of the Chinese Academy of Sciences and IOP Publishing Ltd}%

\begin{multicols}{2}

\section{Introduction}

Resonance is an interesting phenomenon in physics. During the past decades, it has attracted more and more attention in nuclear physics since it plays important roles in understanding many exotic nuclear phenomena, such as the halo~\cite{Tanihataet1985PRL, Meng1996PRL, Poschl1997PRL, Sandulescu2000PRC}, giant halo~\cite{Meng1998PRL, ZhangY2012PRC}, deformed halo~\cite{Hamamoto2010PRC, Zhou2010PRC}, and collective giant resonance~\cite{Curutchet1989PRC, Cao2002PRC}. Therefore, investigation of single-particle resonances has become a hot topic in nuclear physics.

So far, there have been many approaches to study nuclear resonances. They mainly fall into two categories: scattering theory and bound-state-like method. The scattering theory includes the $R$-matrix theory~\cite{Wigner1947PR, Hale1987PRL}, $K$-matrix theory~\cite{Humblet1991PRC}, $S$-matrix method~\cite{Taylor1972NC, Cao2002PRC}, and Jost function approach~\cite{Lu2012PRL, Lu2013PRC}, which have been successfully employed to determine the resonance parameter (energy and width). Recently, the Green's function method~\cite{Economou2006Book, Belyaev1987SJNP, ZhangY2011PRC} has also been demonstrated to be an efficient tool for describing the nuclear single-particle resonant states~\cite{Sun2014PRC}. On the other hand, several bound-state-like methods have also been developed due to their simplicities in computation, including the real stabilization method (RSM)~\cite{Hazi1970PRA, Mandelshtam1993PRL, Kruppa1999PRA, Zhang2008PRC} and the analytic continuation in the coupling constant (ACCC) method~\cite{Kukulin1989Book, Tanaka1997PRC, Tanaka1999PRC, Cattapan2000PRC, Yang2001CPL, Zhang2004PRC, Zhang2012PRC}, and the complex scaling method (CSM)~\cite{Ho1983PRp, Guo2010IJMPE, Feng2010CPL, Liu2013PRA}.

By solving a complex eigenvalue problem, the CSM treats the bound states and resonant states on the same footing. It has been widely used to study resonances not only in atomic nuclei~\cite{Kruppa1997PRL, Arai2006PRC, Myo2012PRC, Liu2012PRC, Shi2014PRC}, but also in the atoms and molecules~\cite{Ho1983PRp, Moiseyev1998PRp}. Recently, the relativistic mean-field (RMF) theory has attracted wide attention due to its successful applications in describing many nuclear structure phenomena~\cite{Ring1996PPNP, Vretenar2005PRp, Meng2006PPNP} and simulating abundances of stellar nucleosynthesis~\cite{Sun2008PRC, Niu2009PRC, Xu2013PRC, Niu2013PLB}. The combination of the complex scaling method with the RMF theory (RMF-CSM) was first developed in Ref.~\cite{Guo2010PRC} and was used to determine the energies and widths of neutron resonant states in $^{120}$Sn, which agree well with the results from the $S$-matrix method, the ACCC method, and the RSM within the framework of the RMF theory (denoted by RMF-S, RMF-ACCC, and RMF-RSM, respectively). Furthermore, the RMF-CSM was employed to investigate the single-neutron resonant states of Zr isotopes and the pseudospin symmetry in the resonant states~\cite{Liu2012APS}.

Unlike the neutron resonant states, the calculations of proton resonant states should take into account the Coulomb field among protons. In the mean-field approximation, there exist the direct (Hartree) and exchange (Fock) terms for the Coulomb field. However, the Coulomb exchange term is usually neglected, since it is very complicated and time consuming due to the nonlocality in the Fock mean field. Recent studies found that the Coulomb exchange term plays an important role in some nuclear phenomena, such as Cabibbo-Kobayash-Maskawa matrix~\cite{Liang2009PRC} and mass differences of mirror nuclei~\cite{Niu2013PRCa}. Some approximation techniques have been developed to effectively take into account the Coulomb exchange term in the Hartree approximation. The local density approximation (LDA) is usually employed to estimate the Coulomb exchange energies in the nonrelativistic framework~\cite{Schnaider1974PLB, Skalski2001PRC, Bloas2011PRC, Anguiano2001NPA}. In Ref.~\cite{Gu2013PRC}, the relativistic local density approximation (RLDA) for the Coulomb exchange term was developed, and it was found that the relativistic corrections are important to reproduce the exact Coulomb exchange energies. Furthermore, guided by the RLDA, a phenomenological formula for the coupling strength of the Coulomb field was proposed in Ref.~\cite{Niu2013PRCa}. The accuracy of this method is even better than the RLDA, and the relative deviations of the Coulomb exchange energy in the calculations with phenomenological formula are less than $1\%$ for Ca, Ni, Sn, and Pb isotopes~\cite{Niu2013PRCa}.

The proton resonant states of $^{120}$Sn has been studied with the RMF-ACCC method~\cite{Zhang2007EPJA}, while the Coulomb exchange term is neglected in the calculations. With the RMF-CSM, the Coulomb exchange term has been taken into account using the phenomenological coupling strength of the Coulomb field~\cite{Zhu2014PRC}. However the influence of the Coulomb exchange term on resonances has not been studied for a isotopic or isotonic chain. Moreover, its influence on the resonance width is not studied as well. Therefore, in this work, we will apply the RMF-CSM to calculate the resonance parameters of Sn isotopes and $N=82$ isotones. The phenomenological formula will be used to evaluate the influence of the Coulomb exchange term on the resonance parameters. Furthermore, the influence of the Coulomb exchange term on single-proton resonance widths will be investigated for the doubly magic nuclei $^{40}$Ca, $^{56,78}$Ni, $^{100,132}$Sn, and $^{208}$Pb.

\section{Numerical details}

The basic ansatz of the RMF theory is a Lagrangian density where nucleons are described as Dirac particles that interact via the exchange of mesons (the scalar $\sigma$, the vector $\omega$, and isovector vector $\rho$) and the photon. From the Lagrangian density, the Dirac equation of the nucleon can be obtained using the classical variation principle. For spherical nuclei, meson fields and densities depend only on the radial coordinate $r$ and the Dirac equation is simplified to the radial Dirac equation, which is
\begin{eqnarray}
  \left(
    \begin{array}{cc}
      V+S+M     & -\frac{d}{dr}-\frac{1}{r}+\frac{\kappa}{r} \\
      \frac{d}{dr}+\frac{1}{r}+\frac{\kappa}{r}      & V-S-M \\
    \end{array}
  \right)
  \left(
    \begin{array}{c}
      f \\
      g \\
    \end{array}
  \right)
  =
  \varepsilon
  \left(
    \begin{array}{c}
      f \\
      g \\
    \end{array}
  \right),
\end{eqnarray}
where $V$ and $S$ are the vector and scalar potentials, and $M$ is the nucleon mass. In this work, those parameters of effective interaction in $V$ and $S$ are taken from the NL3 parameter set~\cite{Lalazissis1997PRC}. To investigate the influence of the Coulomb exchange term, the effective charge factor in Ref.~\cite{Niu2013PRCa},
\begin{eqnarray}\label{Eq:EffCharg}
  \eta(Z,A)= 1 - aZ^b + cA^d,
\end{eqnarray}
is employed with $a=0.366958, b=-0.645775, c=0.030379, d=-0.398341$, which is the factor used to multiply the coupling strength of the Coulomb field.

In the CSM, the Hamiltonian $H$ and wave function $\psi$ are transformed by an unbounded nonunitary scaling operator $U(\theta)$ with a real parameter $\theta$, i.e.,
\begin{eqnarray}
  H \rightarrow H_\theta = U(\theta)HU(\theta)^{-1},
  \quad
  \psi \rightarrow \psi_\theta=U(\theta)\psi.
\end{eqnarray}
The Dirac equation is then transformed to the complex scaled Dirac equation
\begin{eqnarray}\label{Eq:CSDE}
  H_\theta \psi_\theta = \varepsilon \psi_\theta.
\end{eqnarray}
The complex scaled Dirac equation Eq. (\ref{Eq:CSDE}) is solved by expansion in the harmonic oscillator basis as in Ref.~\cite{Guo2010PRC}. The eigenvalues of $H_\theta$ representing continuous spectrum rotate with $\theta$, while eigenvalues representing bound states or resonant states do not change with $\theta$ as long as $\theta$ is large enough to expose them from the continuous spectrum. The latter are associated with resonance complex energies $E_r-i E_i=E_r-i \Gamma/2$, where $E_r$ is resonance energy and $\Gamma$ is the width. For simplicity, the resonance energy and width calculated including the Coulomb exchange term are denoted by $E_r^*$ and $\Gamma^*$, respectively.

\begin{center}
\includegraphics[width=7cm]{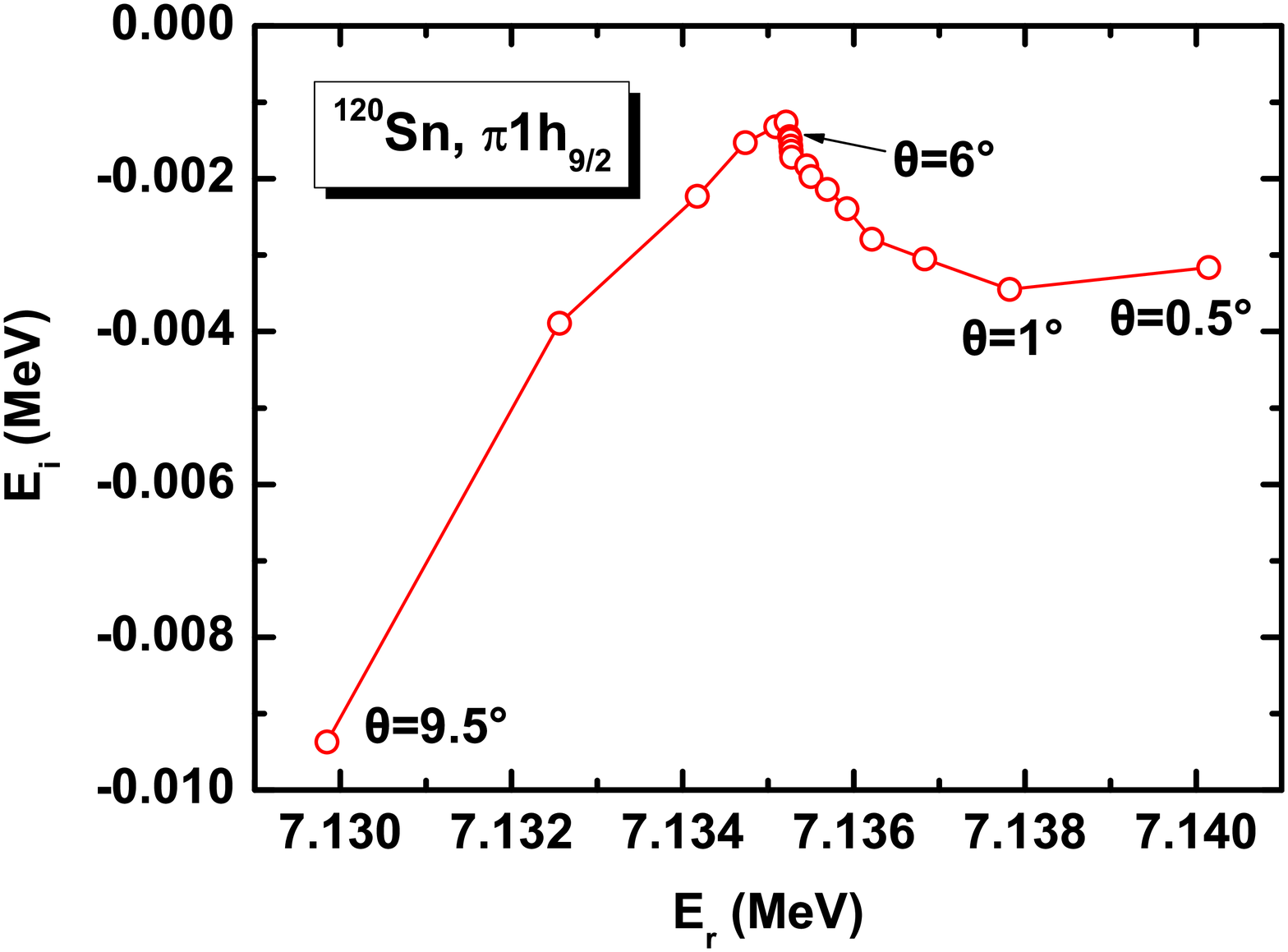}
\figcaption{\label{Fig:ThetaTraj}(Color online) The $\theta$ trajectory of single-proton resonant state 1h$_{9/2}$ for $^{120}$Sn calculated with the RMF-CSM.}
\end{center}

In the actual calculation, there is no resonance energy that is completely independent of $\theta$ due to the numerical approximation. The best estimate for the resonance energy is given by the value of $\theta$ for which the rate of change with respect to $\theta$ is minimal, i.e, the densest place in the $\theta$ trajectory. Fig.~\ref{Fig:ThetaTraj} shows the $\theta$ trajectory of single-proton resonant state $1h_{9/2}$ for $^{120}$Sn calculated using the RMF-CSM. The densest place corresponds to $\theta= 6.0^\circ$ and its energy and width are $7.135$ MeV and $0.0037$ MeV, respectively. The result agrees with the results of the RMF-ACCC and RMF-rS methods (r stands for relativistic and S stands for scattering), whose resonance parameters (energy and width) are ($7.13$ MeV, $0.017$ MeV) and ($7.132$ MeV, $0.003$ MeV), respectively~\cite{Zhang2007EPJA}. This indicates the reliability of the CSM to calculate the parameters of single-proton resonant states.

\section{Results and discussions}

\begin{center}
\includegraphics[width=6cm]{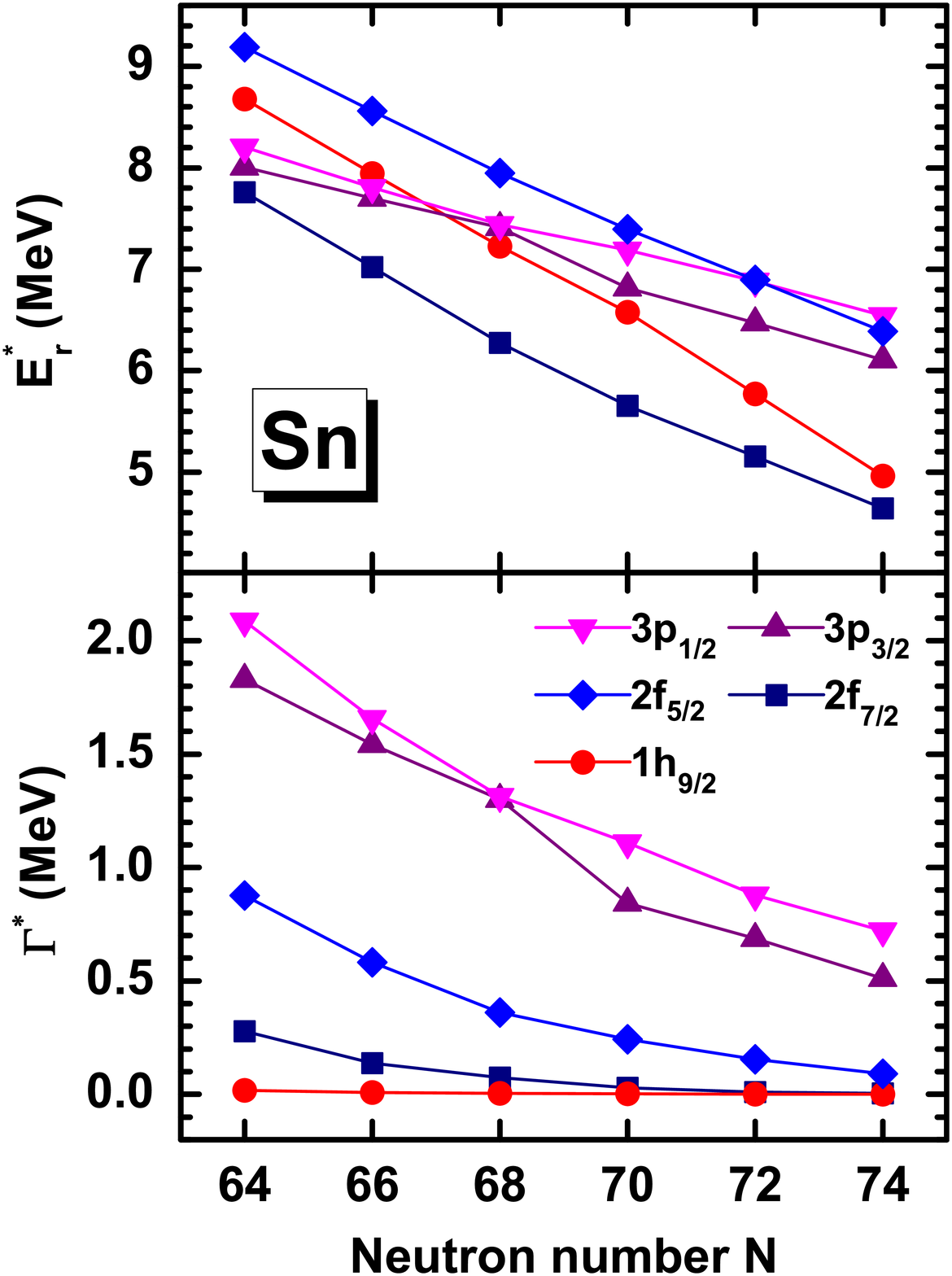}
\figcaption{\label{Fig:ErGamSn}(Color online) Energies $E_r^*$ and widths $\Gamma^*$ of single-proton resonant states for Sn isotopes calculated by the RMF-CSM with the phenomenal Coulomb exchange term.}
\end{center}

With the RMF-CSM, the energies and widths of single-proton resonant states of Sn isotopes are calculated by phenomenologically including the Coulomb exchange term with Eq. (\ref{Eq:EffCharg}). The corresponding results are shown in Fig.~\ref{Fig:ErGamSn}. Clearly, the energies and widths decrease with increasing neutron number. Moreover, the energy decreases more rapidly for the state with larger orbital angular momentum $l$ and hence lead to the crossing phenomenon appearing in the resonance levels. For the spin doublets $(2f_{7/2}, 2f_{5/2})$ and $(3p_{3/2}, 3p_{1/2})$, variation of energy with neutron number is similar due to the same centrifugal potential, while their energy differences are attributed to the spin-orbit potential. This isotopic trend for the resonance energy and width are similar to those without the Coulomb exchange term~\cite{Zhu2014PRC}.

\begin{center}
\includegraphics[width=6cm]{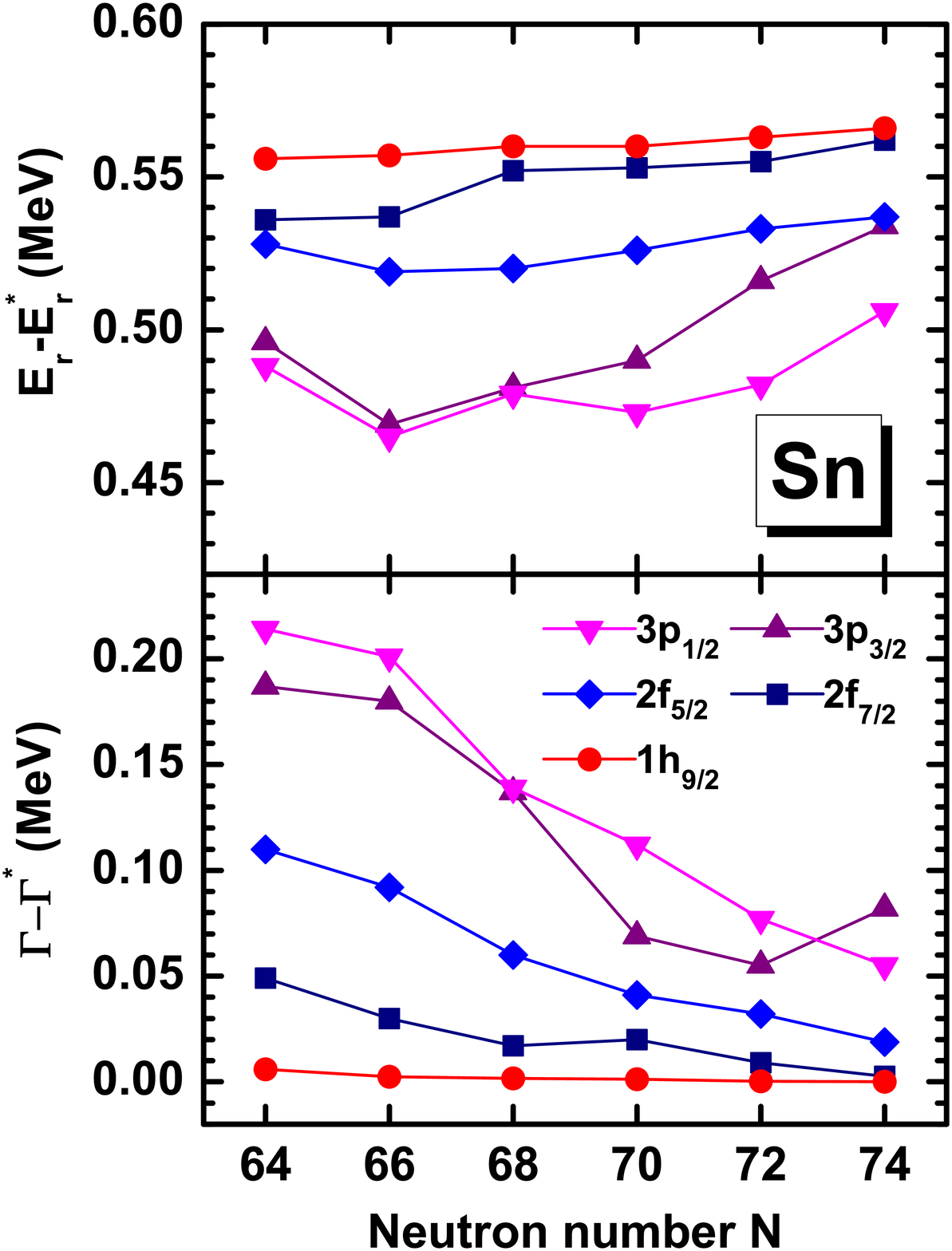}
\figcaption{\label{Fig:DiffErGamSn}(Color online) The differences of single-proton resonance energies ($E_r-E_r^*$) and widths ($\Gamma-\Gamma^*$) for Sn isotopes by phenomenally including the Coulomb exchange term.}
\end{center}

For better understanding the influence of the Coulomb exchange term on the resonance parameters, the differences of resonance energies and widths calculated with and without the Coulomb exchange term are shown in Fig.~\ref{Fig:DiffErGamSn} for Sn isotopes. It is found that the single proton resonance energies and widths of Sn isotopes are reduced when the Coulomb exchange term is phenomenologically taken into account. The difference of the resonance energy caused by the Coulomb exchange term only slightly increases with the increasing neutron number, which is within the range of $0.4-0.6$ MeV, although the resonance energy spans from $4.643$ MeV ($2f_{7/2}$ in $^{124}$Sn) to $9.187$ MeV ($2f_{5/2}$ in $^{114}$Sn). Therefore, the isotopic trend of the resonance energy remains unchanged whether the Coulomb exchange term is taken into account. The difference of the resonance width generally decreases with the increasing neutron number, while its magnitude is relatively small and cannot change the isotopic trend of the resonance width.

To study the influence of the Coulomb exchange term on dependence of resonance parameters on the charge number, the resonance parameters of $N=82$ isotones are calculated by including the Coulomb exchange term phenomenologically. The corresponding results are shown in Fig.~\ref{Fig:ErGamN82}. The Coulomb repulsive force plays an important role in determining the resonance parameters and it is determined by the proton number in nucleus. Therefore, the resonance energy increases with the increasing proton number for $N=82$ isotones and its change is in nearly the same magnitude for various resonant states with different $l$. For the width of resonant state, it also increases with the increasing proton number, while its change is remarkable for broad resonant states, such as $3p_{1/2}$ and $3p_{3/2}$. As the case in Sn isotopes, the change trend of the resonance energy and width for $N=82$ isotones is also similar to that without the Coulomb exchange term~\cite{Zhu2014PRC}.

\begin{center}
\includegraphics[width=6cm]{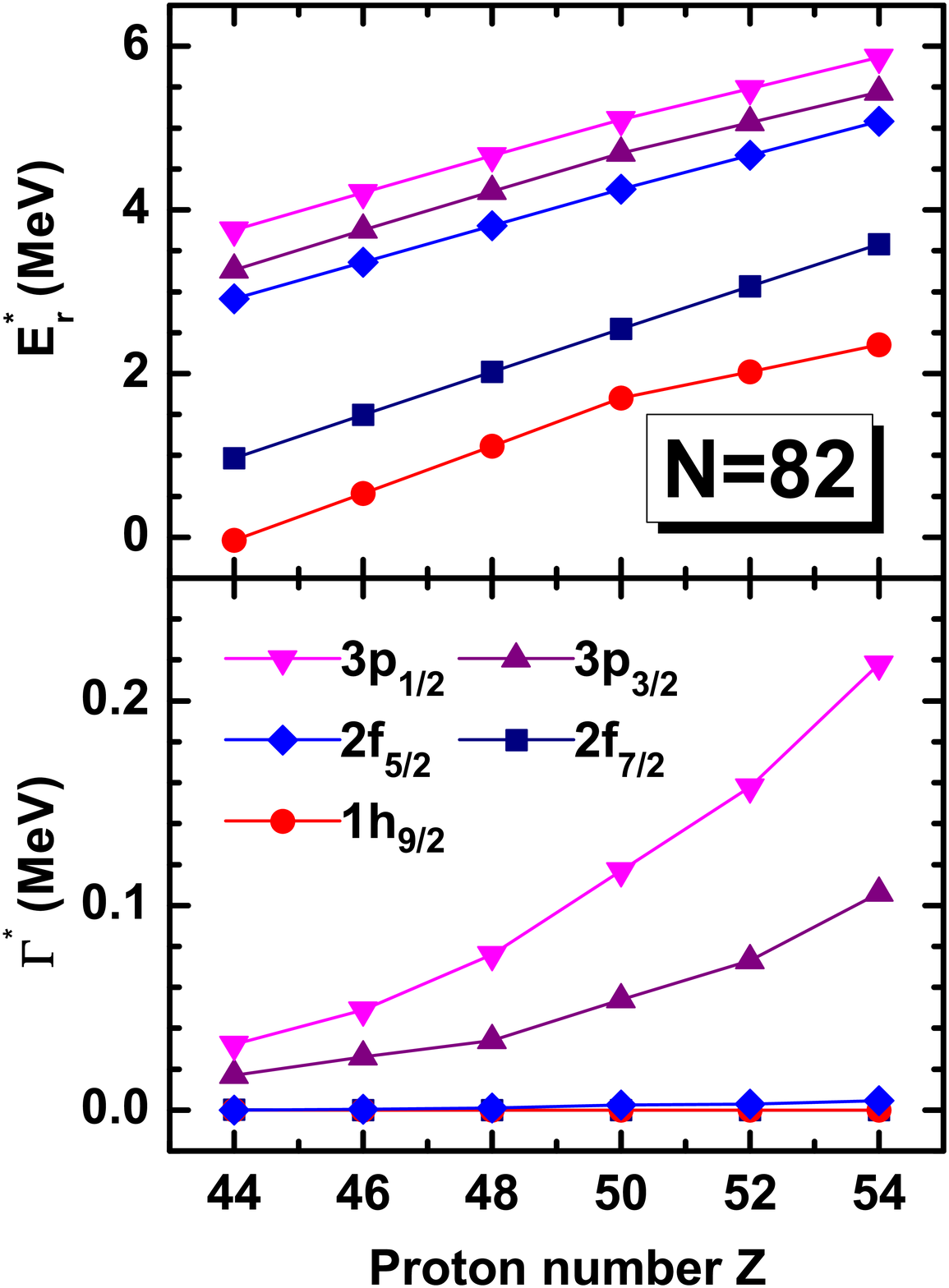}
\figcaption{\label{Fig:ErGamN82}(Color online) Energies $E_r^*$ and widths $\Gamma^*$ of single-proton resonant states for $N=82$ isotones calculated by the RMF-CSM with the Coulomb exchange term.}
\end{center}

In Fig.~\ref{Fig:DiffErGamN82}, the differences of single-proton resonance energies and widths calculated with and without the Coulomb exchange term are shown for $N=82$ isotones. It is found that the resonance energies and widths of $N=82$ isotones are reduced when the coulomb exchange term are taken into account phenomenologically. In general, the difference of the resonance energy caused by the Coulomb exchange term slightly increases with the increasing proton number, and it is still within the range of $0.4-0.6$ MeV as the case in Sn isotopes. Therefore, the isotonic trend of the resonance energy remains unchanged. The difference of the resonance width increases with the increasing proton number, and the change in width even reaches about $0.07$ MeV for the broad resonant states $3p_{1/2}$ and $3p_{3/2}$ of $^{136}$Xe. However, its magnitude is still not large enough to change the trend of the resonance width for $N=82$ isotones.

\begin{center}
\includegraphics[width=6cm]{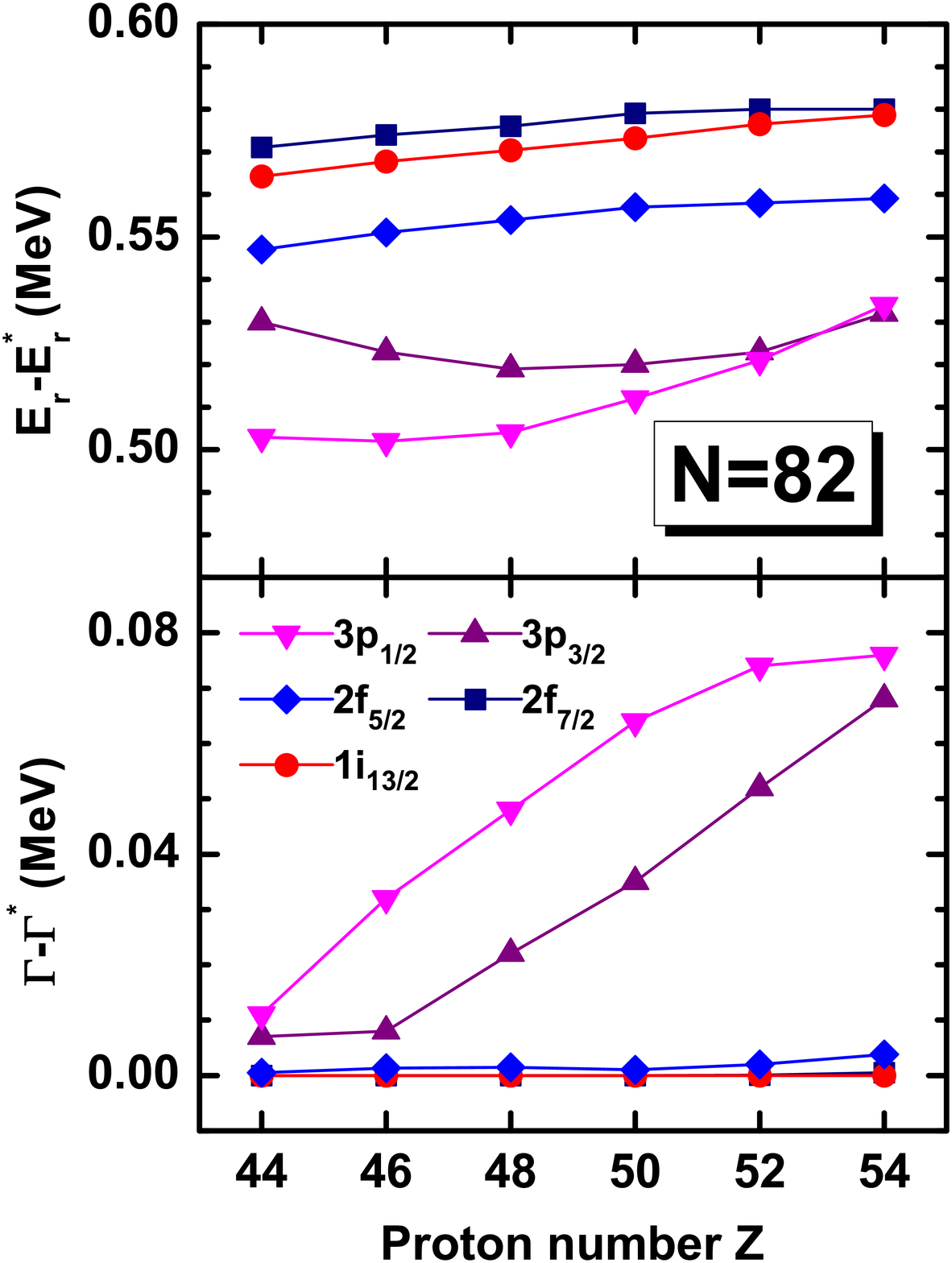}
\figcaption{\label{Fig:DiffErGamN82}(Color online) The differences of single-proton resonance energies ($E_r-E_r^*$) and widths ($\Gamma-\Gamma^*$) for $N=82$ isotones by phenomenally including the Coulomb exchange term.}
\end{center}

\begin{center}
\includegraphics[width=7cm]{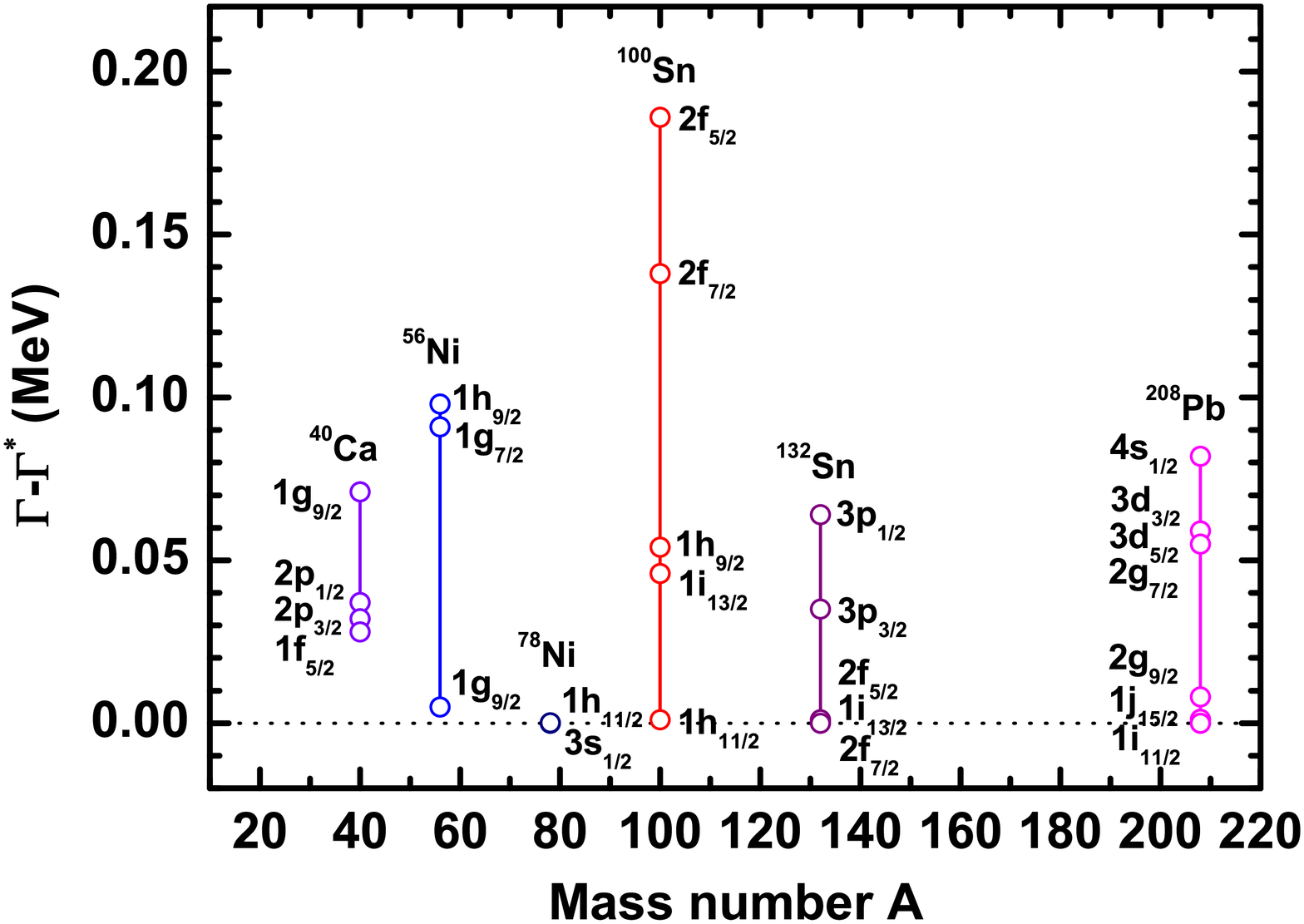}
\figcaption{\label{Fig:DiffGamMagNucl}The difference of single-proton resonance widths for doubly magic nuclei $^{40}$Ca, $^{56,78}$Ni, $^{100,132}$Sn, and $^{208}$Pb by phenomenally including the Coulomb exchange term.}
\end{center}

Through the above research, it is found that the Coulomb exchange term plays an important role not only in the resonance energy, but also in resonance width. In Ref.~\cite{Zhu2014PRC}, the influence of the Coulomb exchange term on single-proton resonance energy has been studied in different nuclear region, taking the doubly magic nuclei $^{40}$Ca, $^{56,78}$Ni, $^{100,132}$Sn, and $^{208}$Pb as examples. So it is interesting to further investigate the influence of the Coulomb exchange term on the single-proton resonance width. In Fig.~\ref{Fig:DiffGamMagNucl}, the differences of single-proton resonance widths for doubly magic nuclei $^{40}$Ca, $^{56,78}$Ni, $^{100,132}$Sn, and $^{208}$Pb are shown. It is clear that the Coulomb exchange term reduces the resonance width, whose reduction is within $0.2$ MeV. In addition, the change of the width is larger for the spin unaligned state than that for the spin aligned state, such as the spin doublets ($3p_{1/2}$, $3p_{3/2}$) of $^{40}$Ca. In an isotope, the state with the same quantum number always decreases with the increasing neutron number, such as $2f_{5/2}$ in $^{100, 132}$Sn. These indicate that the contribution of the Coulomb exchange term to the proton resonance width is related to the specific nucleus and the quantum numbers of the state.

\section{Summary}

In summary, the influence of the Coulomb exchange term on the single-proton resonances is investigated with the complex scaling method based on the relativistic mean-field model. Taking Sn isotopes and $N=82$ isotones as examples, it is found that the Coulomb exchange term reduces the single-proton resonance energy, and the dependence of this influence on neutron number and proton number is so weak that energy reduction is within the range of $0.4-0.6$ MeV. Therefore, the isotopic and isotonic trends of the resonance energy are similar to those without the Coulomb exchange term. Moreover, the Coulomb exchange term also reduce the single-proton resonance width, and the reduction is generally larger for the border single-proton resonance. For Sn isotopes, the influence of the Coulomb exchange term on the resonance width generally decreases with the increasing neutron number, while it increases with the increasing proton number for $N=82$ isotones. However, the influence of the Coulomb exchange term on resonance width is still not large enough to change the trend of resonance width with respect to the neutron number and proton number. Furthermore, the influence of the Coulomb exchange term on resonance width is further investigated in a wide range of nuclear region, taking the doubly magic nuclei $^{40}$Ca, $^{56,78}$Ni, $^{100,132}$Sn, and $^{208}$Pb as examples. It is found that the Coulomb exchange term reduces the proton resonance width within $0.2$ MeV, whose magnitude depends on the specific nucleus and the quantum numbers of resonant states.

\vspace{3mm}
\acknowledgments{Helpful discussions with GUO Jian-You and SHI Min are acknowledged.}

\end{multicols}

\clearpage

%\end{CJK*}
\end{document}